\documentclass[12pt]{article}
\usepackage[latin9]{inputenc}
\usepackage{amsmath}
\usepackage{amssymb,color}
\usepackage{graphicx}
\usepackage{esint}

\makeatletter
\usepackage{subfigure}\usepackage{amscd}
\usepackage{appendix}
\usepackage{verbatim}
\usepackage{epstopdf}

\makeatother

\begin{document}
\begin{titlepage}

{\hbox to\hsize{\hfill{}April 2018 }}

\bigskip{}
\vspace{3\baselineskip}

\begin{center}
\textbf{\Large{}Exotic Lepton Searches via Bound State Production
at the LHC}
\par\end{center}{\Large \par}

\begin{center}
\textbf{Neil D. Barrie$^{1,2}$, Archil Kobakhidze$^{1}$, Shelley Liang$^{1}$,\\
Matthew Talia$^{1}$ and Lei Wu$^{3}$ }\\
\textbf{ }
\par\end{center}

\begin{center}
{ \textit{\small{}$^{1}$ ARC Centre of Excellence for Particle Physics
at the Terascale, }\\
\textit{\small{} School of Physics, The University of Sydney, NSW
2006, Australia }\\
 \textit{\small{}$^{2}$ Kavli IPMU (WPI), UTIAS, University of Tokyo, Kashiwa, Chiba 277-8583, Japan }\\
\textit{\small{} $^{3}$ Department of Physics and Institute of Theoretical
Physics, Nanjing Normal University, }\\
\textit{\small{} Nanjing, Jiangsu 210023, China}\\
\textit{\small{} E-mails: neil.barrie, archil.kobakhidze, shelley.liang,
matthew.talia, lei.wu1@sydney.edu.au }\\
\textit{\small{}}}
\par\end{center}{\small \par}

\begin{center}
\textbf{\large{}Abstract}
\par\end{center}{\large \par}

\noindent Heavy long-lived multi-charged leptons (MCLs) are predicted
by various new physics models. These hypothetical MCLs can form bound
states, due to their high electric charges and long life times. In this work, we propose a novel strategy of searching for MCLs through their bound state productions and decays. By
utilizing LHC-8 TeV data in searching for resonances in the diphoton
channel, we exclude the masses of isospin singlet heavy leptons with
electric charge $|q|\geq 6$ (in units of electron charge) lower than
$\sim$1.2 TeV, which are much stronger than the corresponding 8 TeV LHC bounds from analysing the high ionisation and the long time-of-flight
of MCLs. By utilising the current 13 TeV LHC diphoton channel measurements the bound can further exclude MCL masses up to  $\sim$1.6 TeV for $|q|\geq 6$.
Also, we demonstrate that the conventional LHC limits from searching for MCLs produced via Drell-Yan processes can be enhanced by including the contribution of photon fusion processes.

\end{titlepage}

\section{Introduction}

Heavy long-lived multi-charged leptons (MCLs) are predicted by various
extensions of the Standard Model (SM) (for a review, see \cite{Fairbairn20071}).
The charge conservation of such MCLs typically implies that they are
long-lived. Theoretically, the observed charge quantisation of known
quarks and leptons lacks a fundamental explanation within the SM due
to the Abelian nature of the hypercharge gauge symmetry. Various theoretical
frameworks have been proposed to accommodate charge quantisation,
such as quantum-mechanical monopoles \cite{Dirac:1931kp}, grand unified
theories \cite{Georgi:1974sy} and gauge anomaly cancellation \cite{Foot:1992ui}.
Experimental observation of non-integer and multiply charged particles
may thus have important implications for the charge quantisation problem
and beyond the Standard Model physics in general.

In LHC experiments, MCLs with lifetimes greater than ${\cal O}(1)$
ns can be observed with the ATLAS and CMS detectors as high-momentum
tracks with anomalously large rates of energy loss through ionization.
MCLs could also be highly penetrating so that the fraction reaching
the muon system of the detectors would be sizeable. Therefore, the
muon system could be used to help in identification and in the measurement
of the time-of-flight (TOF) of the particles. So far, the ATLAS and
CMS collaborations have extensively searched for long-lived MCLs by
analysing the anomalously high ionisation and the long TOF to the
outer muon system at the LHC \cite{cms-multicharged,Chatrchyan:2013oca,atlas-multicharged}.
Based on the 8 TeV dataset, considering Drell-Yan (DY) like signals, the ATLAS collaboration has excluded the
MCLs mass range from 50 GeV up to 660, 740, 780, 785, and 760 GeV
for integer charges $|q|=2-6$, respectively \cite{atlas-multicharged}.
 Similar bounds are obtained by the CMS collaboration.
In this case,  integer charges $|q|=1-8$
are excluded for masses below 574, 685, 752, 793, 796, 781, 757, and
715 GeV/c$^{2}$, respectively \cite{Chatrchyan:2013oca}. The non-integer
charge analysis for certain charge ranges and masses are also given by the CMS
collaboration in \cite{cms-multicharged}.

However, the signal efficiency in searches for the MCLs drops significantly
with the increase of the mass and charge of leptons. For example,
when the charge is higher than 6, the signal efficiency is expected
to be less than 5\% \cite{atlas-multicharged}. Such low efficiencies
mean that a different approach is required.

Due to their high charges and long life times, the MCLs are expected
to form bound states via the electromagnetic interaction, which will
subsequently decay to SM particles. We call these bound states `leptonium'.
In this paper we propose a new search for the MCLs through their bound
state productions and decays at the LHC. To demonstrate our method, we consider vector-like,
weak isospin singlet leptons $L$, charged only under $U(1)_{Y}$
with hypercharge (equal to the electric charge) $Y_{L}$. Our proposal
can also be applied to other representations of MCLs with modifications
in the production and decay calculations.

\section{Constraints on MCLs}
\begin{figure}[h]
	\centering
	\subfigure{\includegraphics[width=0.49\textwidth]{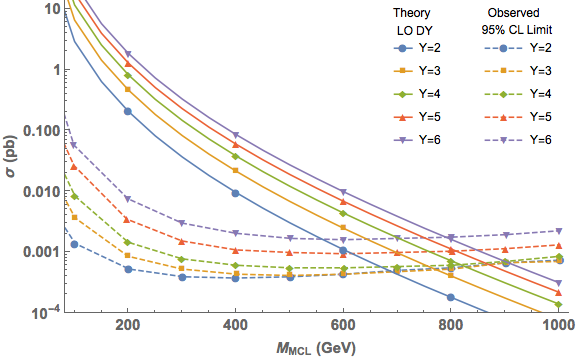}}
	\subfigure{\includegraphics[width=0.49\textwidth]{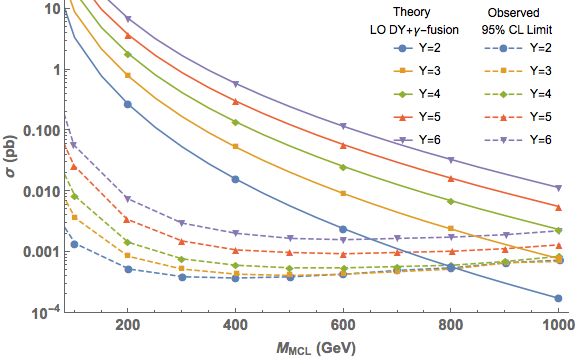}}
\caption{{\small{}} LHC-8 TeV exclusion limits of MCLs from DY process (left panel) and from DY+photon fusion process (right panel). The LHC-8 TeV 95\% C.L. observed limits (dashed lines) are taken from \cite{atlas-multicharged}.}
\label{pair} \end{figure}

In current LHC experiment of searches for MCLs~\cite{cms-multicharged,Chatrchyan:2013oca,atlas-multicharged}, only pair production of MCLs via DY processes is included. However, these MCLs can also be produced in pairs via the photon fusion process. In Fig.~\ref{pair}, we compare the LHC-8 TeV exclusion limits of MCLs from DY processes (left panel) with those from DY+photon fusion processes (right panel). We compute these processes using \texttt{Madgraph 5} \cite{Alwall:2014hca} with the \texttt{NNPDF2.3QED} parton distribution function (PDF) \cite{Ball2013}. We can see that the DY process alone can exclude the masses of the heavy charged lepton up to 640 and 800 GeV for $|q|=2$ and $3$ respectively \footnote{Since the 95\% C.L. observed limits on MCLs are within 1 TeV in ATLAS analysis at 8 TeV LHC, we will not explicitly state the comparison between DY and DY+photon fusion for charges $|q|=4-6$.}. While the DY+photon fusion process can exclude masses up to 810 and 1000 GeV for $|q|=2$ and $3$ respectively. Therefore, we can conclude that the contribution of the photon fusion process to LHC searches for MCLs is not negligible and should be included in LHC analyses.

Since the multi-charged lepton can contribute to the electroweak observables,
we consider the dependence of the $\Delta S,\Delta T,\Delta U$ parameters
\cite{Peskin:1991sw} on the hypercharge and mass of the new lepton.
\begin{equation}
\Delta S=\frac{4}{3\pi}\frac{Y_{L}^{2}\sin^{4}\theta_{w}}{1-4m_{L}^{2}/m_{Z}^{2}},~
\Delta T=\frac{1}{3\pi}\frac{Y_{L}^{2}\tan^{2}\theta_{w}}{1-4m_{L}^{2}/m_{Z}^{2}},~
\Delta U=\frac{1}{3\pi}\sin^{4}\theta_{w}\frac{m_{Z}^{2}}{m_{L}^{2}}\left(Y_{L}^{2}+\frac{2}{3\sin^{2}\theta_{w}}\right)
\end{equation}
comparing with the experimental values ($\pm1\sigma$) are \cite{Baak:2014ora}:
\begin{equation}
\Delta S=0.05\pm0.11,~~\Delta T=0.09\pm0.13,~~\Delta U=0.01\pm0.11
\end{equation}
we find that the $T$ parameter is the most constraining, and require:

\begin{equation}
|Y_{L}|<9\left(\frac{m_{L}}{300~\textrm{GeV}}\right),
\end{equation}
in the limit $m_{L}\gg m_{Z}$. Hence for MCLs masses above $\sim300$ GeV with $ Y_{L}\leq 9 $, the oblique parameter bounds can be easily satisfied.

The Millikan-type experiments require the abundance of fractionally
charged particles be less than $\sim10^{-22}$ per ordinary matter
nucleon \cite{Perl:2004qc}. However, this bound can be easily satisfied
by assuming a low reheating temperature after inflation, $T_{rh}<m_{L}$ \cite{Giudice:2000ex}.

The introduction of these leptons would also be expected to have implications
for the running of the hypercharge coupling. It is found that the
large hypercharge $Y_{L}$ accelerates the running of the effective
hypercharge coupling constant such that it hits the Landau pole at
the following energy scale:
\begin{equation}
\Lambda^{2}=m_{Z}^{2}\exp\left(\frac{1}{8+Y_{L}^{2}}\frac{3\pi}{\alpha(m_{Z}^{2})}\right)
\end{equation}
which gives $\sim10^{5}-10^{17}$ GeV for hypercharges $Y_{L}=8-3$
at the TeV mass scale, respectively. Although this shows that our
perturbative calculations in the energy domain well below the Landau
pole are valid, the hypercharge $U(1)$ should probably be embedded
into a larger non-Abelian group in order to avoid theoretical inconsistency.
Alternatively, the hypercharge $U(1)$ may avoid the Landau pole by
developing an ultraviolet non-Gaussian fixed point \cite{Holdom:2010qs}.

\section{MCL Bound State Production at the LHC}

Besides the production of free MCLs, these hypothetical MCLs can form bound
states via the electromagnetic interaction, due to their high electric
charges and long life times. Such heavy leptoniums can be copiously
produced at the LHC and serve as a smoking gun to probe the MCLs.

In the non-relativistic approximation, the heavy lepton $L$ is described
by the Schr\"odinger equation
\begin{equation}
\left(-\frac{\nabla^{2}}{m_{L}}+V(r)\right)\psi=E\psi~,\label{2.1}
\end{equation}
with the binding Coulomb potential
\begin{equation}
V(r)=-\frac{Y_{L}^{2}\alpha}{r}~,\label{2.2}
\end{equation}
where $\alpha\approx1/128$ is the fine structure constant evaluated
at $m_{Z}$ \footnote{Note that since the leptonium Bohr radius, $r_{L}=\frac{2}{m_{L}}\frac{1}{Y_{L}^{2}\alpha}$,
is larger than the Compton wavelength of the $Z$-boson, $r_{L}\gtrsim1/m_{Z}$,
in deriving the potential (\ref{2.2}) we only take into account massless
photon exchange.}. In this approximation, the ground state ($n=1,~l=0$) energy is
given by:
\begin{equation}
E=-\frac{1}{4}m_{L}\left(Y_{L}^{2}\alpha\right)^{2}~,\label{2.3}
\end{equation}
We also include the leading $\left[\sim\mathcal{O}\left(Y_{L}^{8}\alpha^{4}\right)\right]$
relativistic Breit correction to the binding energy given in Eq. (\ref{2.3}):
\begin{equation}
\delta E_{{\rm Breit}}=-\frac{1}{2m_{L}}\left(E^{2}-2E\left\langle V\right\rangle +\left\langle V^{2}\right\rangle \right)=-\frac{5}{16}m_{L}\left(Y_{L}^{2}\alpha\right)^{4}~.\label{2.4}
\end{equation}
The mass of the para-leptonium $\psi_{para}$ ($J^{PC}=0^{-+}$) is
then given by:
\begin{equation}
m_{\psi_{para}}=2m_{L}+E+\delta E_{{\rm Breit}},\label{2.5}
\end{equation}
and from this we can derive the mass of the constituent leptons from
a resonance observed in the diphoton channel which can be identified
as the para-leptonium state,
\begin{equation}
m_{L}=m_{\psi_{para}}\left(2-\frac{1}{4}(\alpha Y_{L}^{2})^{2}-\frac{5}{16}(\alpha Y_{L}^{2})^{4}\right)^{-1}~.\label{lepton_mass}
\end{equation}

The wave function $\psi_{para}(0)$ is given by,
\begin{equation}
\left|\psi_{para}(0)\right|^{2}=\left(\frac{1}{\sqrt{4\pi}}\right)^{2}\left|R_{para}(0)\right|^{2},\label{2.8}
\end{equation}
with the radial part evaluated as:
\begin{equation}
\left|R_{para}(0)\right|^{2}=\frac{\left(Y_{L}^{2}\alpha m_{L}\right)^{3}}{2}.\label{2.8b}
\end{equation}
In addition to para-leptonium, our model predicts a heavier spin-1
ortho-leptonium bound state $\psi_{ortho}$ ($J^{PC}=1^{--}$) with
mass \cite{Efimov:2010ih},
\begin{equation}
m_{\psi_{ortho}}\simeq m_{\psi_{para}}\left(1+\frac{7}{12}\frac{\left(Y_{L}^{2}\alpha\right)^{4}}{\left(2-\frac{1}{4}\left(Y_{L}^{2}\alpha\right)^{2}-\frac{5}{16}\left(Y_{L}^{2}\alpha\right)^{4}\right)}\right),\label{ortho}
\end{equation}
The wave function $\psi_{ortho}(0)$ is equal to $\psi_{para}(0)$
at leading order, since they satisfy the same Schr\''oinger equation.

\begin{figure}[ht]
\centering \includegraphics[width=1\textwidth]{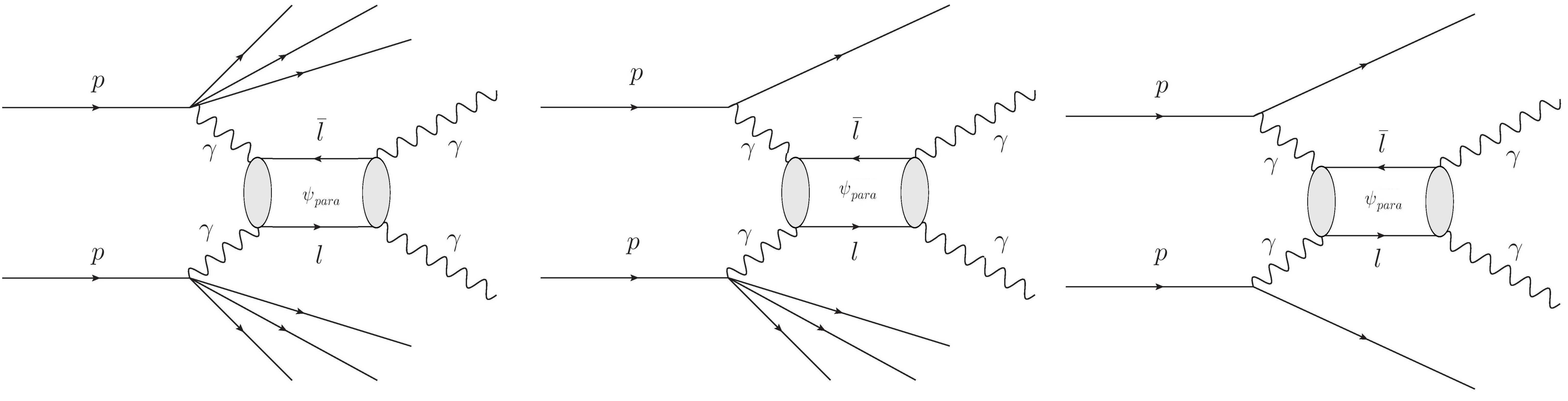} \caption{{\small{}The Feynman diagrams for inelastic (left), semi-elastic (middle) and elastic (right) scattering photoproduction subprocesses.}}
\label{feyn}
\end{figure}

At the LHC the spin-0 para-leptonium is predominantly produced via
photon-fusion processes in three distinct ways, shown in Fig.~\ref{feyn}.
Photoproduction is dominated by inelastic scattering, which is followed
by the semi-elastic and elastic processes in the ratio 63:33:4 ~\cite{csaki2016}. Here it should be noted that this relation  changes only slightly with the heavy lepton mass since its dependence on the mass is cancelled in the ratio~\cite{Luszczak2015}. So we use this ratio as a good approximation in the numerical calculations that follow. The
parton level cross section for photoproduction of the bound state
$\psi_{para}$ can be written in terms of the decay width of $\psi_{para}\to\gamma\gamma$,
since both the production and decay processes share the same matrix
elements. This is given by \cite{Bertulani:2001zk,Kats:2012ym}:
\begin{equation}
\hat{\sigma}_{\gamma\gamma\rightarrow\psi_{para}}(\hat{s})=8\pi^{2}\frac{\Gamma_{\psi_{para}\rightarrow\gamma\gamma}}{m_{\psi_{para}}}\delta(\hat{s}-m_{\psi_{para}}^{2})\label{2.7}
\end{equation}
The decay width $\Gamma_{\psi_{para}\rightarrow\gamma\gamma}$ in
turn is given in terms of the annihilation cross section of a free
lepton-antilepton pair into two photons and the wave function for
the leptonium bound state evaluated at the origin:
\begin{equation}
\Gamma_{\psi_{para}\rightarrow\gamma\gamma}=\frac{16\pi\alpha^{2}Y_{L}^{4}\left|\psi_{para}(0)\right|^{2}}{m_{\psi_{para}}^{2}}~.
\end{equation}
Then we can calculate the two photon production and decay cross section
by convolution with the parton distribution function (PDF) for the
photon in the proton, $f_{\gamma}(x)$:
\begin{equation}
\sigma_{\gamma\gamma\rightarrow\psi_{para}\rightarrow\gamma\gamma}=\frac{8\pi^{2}}{sm_{\psi_{para}}}\frac{\Gamma_{\psi_{para}\rightarrow\gamma\gamma}^{2}}{\Gamma_{\psi_{para}}}\int\delta(x_{1}x_{2}-m_{\psi_{para}}^{2}/s)f_{\gamma}(x_{1})f_{\gamma}(x_{2})dx_{1}dx_{2}\label{2.9}
\end{equation}
where $\sqrt{s}$ is the centre-of-mass energy,
and $\Gamma_{\psi_{para}}$ denotes the leptonium
total decay width. It should be noted that the para-leptonium state
would also decay into $\gamma Z$ and $ZZ$
 , the ratio of these signals would be $1:2\tan^{2}\theta_{W}:\tan^{4}\theta_{W}$,
for $\gamma\gamma:\gamma Z:ZZ$ respectively ($\theta_{W}$ is the weak mixing angle, $\tan\theta_{W}\approx0.55$). Given the sizeable branching
ratio  and clean signatures in the diphoton channel, we use the LHC data of diphoton resonance
searches to obtain bounds on MCLs. Other searches
 in the $Z\gamma$ and $ZZ$ channels, are suppressed by the subsequent decay branching ratios of the $ Z $. In the numerical calculations, we use the \texttt{NNPDF2.3QED}
photon PDF  to calculate the hadronic cross section at the LHC.

\begin{figure}[h]
\centering
\subfigure{\includegraphics[width=0.49\textwidth]{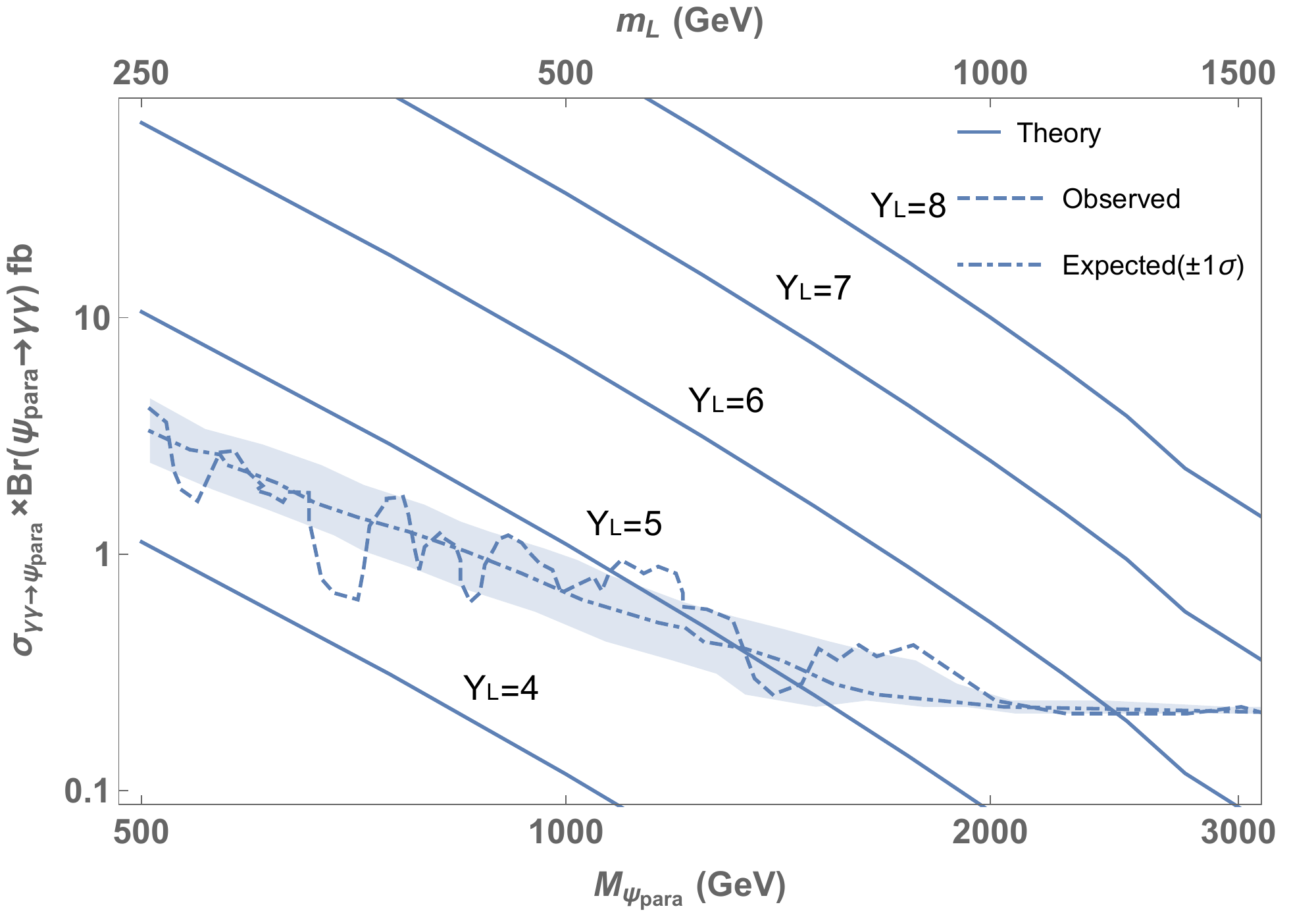}}
\subfigure{\includegraphics[width=0.49\textwidth]{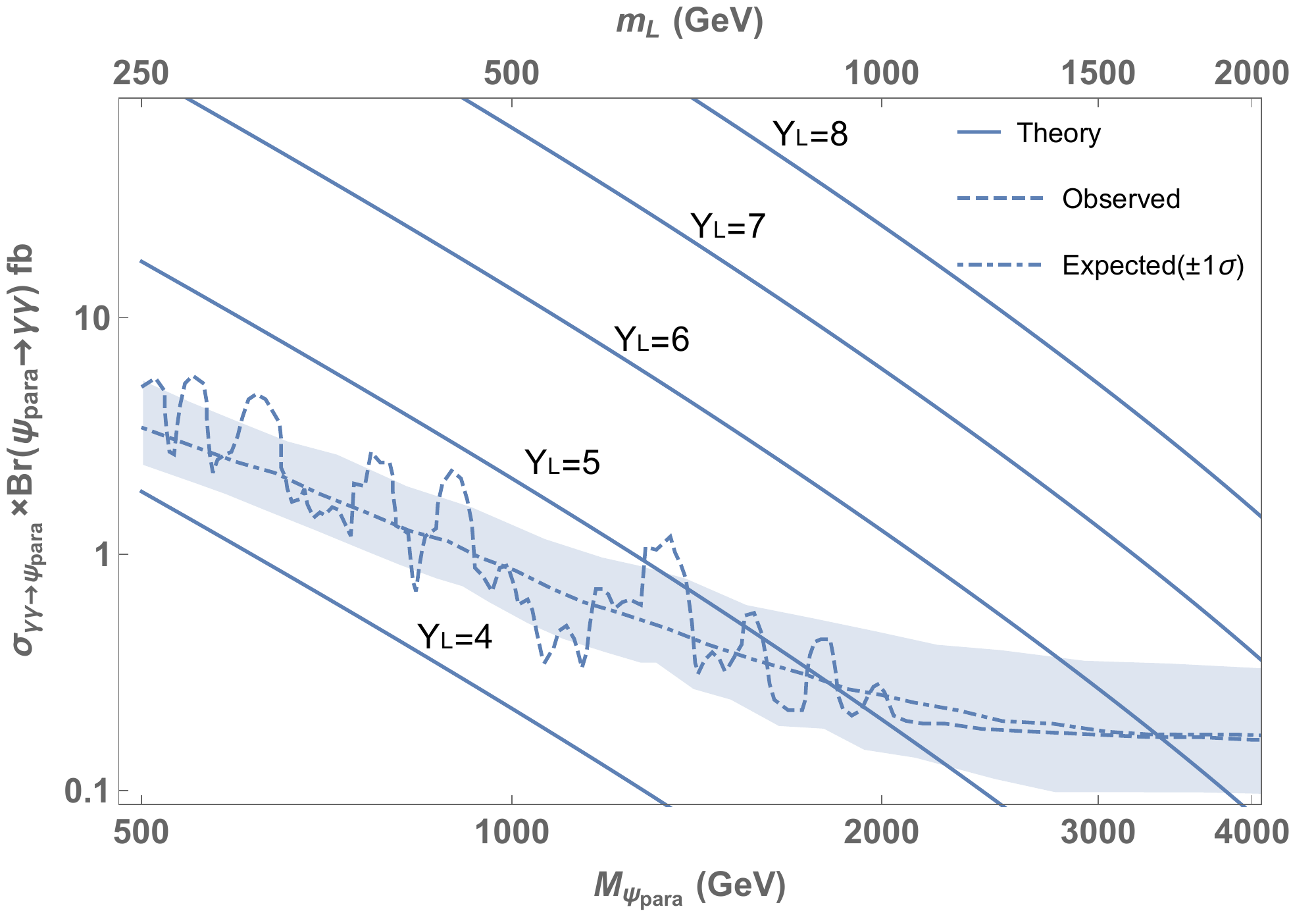}}
\caption{{\small{}The cross section of para-leptonium $\psi_{para}$ decaying
to two photons at LHC-8 TeV (left) and LHC-13 TeV (right) (solid lines). The lower and upper x-axes
represent the masses of para-leptonium and constituent lepton, respectively.
The 95\% CL upper limits are taken from CMS searches \cite{CMS:2015cwa,Khachatryan:2016yec}.}}
\label{plot_diphoton}
\end{figure}

In Fig.~\ref{plot_diphoton}, we plot the dependence of the production
rate of para-leptonium diphoton decay on the masses of $m_{\psi_{para}}$
and $m_{L}$ for integer hypercharges $Y_{L}=4-8$. We can see that 8 TeV LHC data searching for resonances in the diphoton channel can exclude the mass of a heavy lepton with hypercharge $Y_{L}=6$ and $ 7 $ up to 1.2 and 1.5 TeV respectively, which is much stronger than the corresponding ATLAS and CMS bound from analysing the high ionisation and the long time-of-flight in DY production. Also we checked and found that even with the inclusion of the photon fusion process given in Fig.~\ref{pair}, the bounds derived from the diphoton channel are still stronger. The LHC-13 TeV data for the diphoton channel further excludes the mass of an MCL with $Y_{L}\geq6$ below $\sim$ 1.7 TeV at $95\%$ C.L..

While we have concentrated on the lowest energy bound state, a tower of multiple higher energy excited states is also present. The energy difference between these resonances $\Delta E$ are small enough to be resolved at the LHC. Therefore, in a more accurate treatment, interference between multi-resonance amplitudes must be taken into account. However, we have verified that the width of the lightest para-leptonium resonance is much smaller than $\Delta E$, $\Gamma_{\psi_{para}}/\Delta E \propto (\alpha Y^2_L)^3 \ll 1$ for the $Y_L$ range considered, meaning that the interference effects with heavier resonances can be safely ignored for the purpose of obtaining the conservative bounds.

Besides, a spin-1 ortho-leptonium bound state $\psi_{ortho}$ ($J^{PC}=1^{--}$) can be produced via quark-antiquark annihilation at the LHC and may decay into $W^{+}W^{-}$, $f\bar{f}$ ($f=e,\mu,\tau,u,d,c,s,b,t$) or $3\gamma$ final states. The interference effects between higher resonance states for large $Y_L$ may also affect the observation of $\psi_{ortho}$ at the LHC, which requires a more detailed analysis and is beyond the scope of this work.

\section{Conclusion}

Due to their high electric charges and long lifetimes, the heavy exotic
leptons can form bound states, which can be copiously produced at
the LHC. In this paper, we have proposed a new method using the LHC data of searches
for heavy resonances to probe these heavy long-lived multi-charged leptons. The bounds derived from the 8 TeV LHC diphoton results are much stronger than the currently available 8 TeV LHC limits from analysing the high
ionisation and long time-of-flight of freely produced exotic leptons. Furthermore, the mass of an isospin singlet heavy lepton with electric
charge $Y_{L}\geq6$ can be excluded below $\sim$1.7 TeV from diphoton channel 13 TeV data. Therefore, our proposal will prove to be an invaluable new tool in the search for MCLs in the future LHC experiment.

\paragraph{Acknowledgement.}

This work is partially supported by the Australian Research Council. NDB is supported in part by World Premier International Research Center Initiative (WPI), MEXT, Japan. LW is also supported in part by the National Natural Science
Foundation of China (NNSFC) under grant No. 11705093.

\end{document}